\documentclass[aps,prb,showpacs,preprintnumbers,amsmath,amssymb,psfrac,twocolumn]{revtex4}
\usepackage{epsfig}
\usepackage{graphicx}
\usepackage{dcolumn}
\usepackage{bm}

\newcommand{\alphai}{\alpha_i^{{\vk},k_\perp}}
\newcommand{\alphaj}{\alpha_j^{{\vk},k_\perp}}
\newcommand{\alphaip}{\alpha_i^{{\vk}',k_\perp'}}
\newcommand{\alphajp}{\alpha_j^{{\vk}',k_\perp'}}
\newcommand{\vk}{{\bf \vec {k}}_\parallel}

\newcommand{\vf}{v_{\rm F}}

\newcommand{\ef}{\epsilon_{\rm F}}

\newcommand{\tp}{t_\perp}

\newcommand{\mk}{| {\bf \vec{k}}_\parallel |}

\newcommand{\kp}{k_\perp}
\newcommand{\kpp}{k_\perp'}

\newcommand{\eo}{\epsilon_0}
\newcommand{\aAM}{\alpha_{{\vk} , \kp^n , M}}
\newcommand{\aBM}{\beta_{{\vk} , \kp^n , M}}
\newcommand{\aA}{\alpha_{{\vk} , \kp^n }}
\newcommand{\aB}{\beta_{{\vk} , \kp^n }}
\newcommand{\chiintra}{\chi^{\rm intra}}
\newcommand{\chiinter}{\chi^{\rm inter}}
\newcommand{\chitintra}{\tilde{\chi}^{\rm intra}}
\newcommand{\chitinter}{\tilde{\chi}^{\rm inter}}
\begin{document}

\title{Charge distribution and screening in layered graphene systems.}
\author{F. Guinea}

\affiliation{Instituto de Ciencia de Materiales de
  Madrid. CSIC. Cantoblanco. 28049 Madrid. Spain}

\begin{abstract}
The charge distribution induced by external fields in finite stacks of
graphene planes, or in semiinfinite graphite is considered. The interlayer
electronic hybridization is described by a nearest neighbor hopping term, and
the charge induced by the self consistent electrostatic potential is
calculated within linear response (RPA). The screening properties are
determined by contributions from inter- and intraband electronic
transitions. In neutral systems, only interband transitions contribute to the
charge polarizability, leading to insulating-like screening properties, and
to oscillations in the induced charge, with a period equal to the interlayer
spacing. In doped systems, we find a screening length equivalent to 2-3
graphene layers, superimposed to significant charge oscillations.
\end{abstract}
\pacs{73.21.-b; 73.22.-f; 73.21.Ac; 73.20.-r; 73.20.Hb; 73.23.-b; 73.43.-f}

\maketitle
\section{Introduction.}
Electronically charged systems made up of a few graphene layers are being
intensively
studied\cite{Betal04,Netal05b,Netal05,Zetal05b,Zetal05,Betal05,Zetal05c,Betal06},
as well as the
effects of charge accumulation on the surface of bulk
graphite\cite{Metal05,Yetal06}. Hence, the study of the charge distribution
in finite stacks of graphene layers, or in semiinfinite graphite, is a
problem of current interest. 

It is known that screening in a single graphene layer shows anomalous
properties, due to the vanishing of the density of states at the Fermi level
in neutral graphene\cite{GGV97b,GGV99}. A stack of graphene planes where
electrons are confined to each layer also shows unusual screening
properties\cite{GGV01}. In addition, defects and edges can lead to self doping
effects in a single graphene layer\cite{PGN06}. 

The screening properties of bulk graphite were initially investigated
describing the system as a stack of two dimensional
electron fluids electrostaticslly coupled\cite{VF71}. The electronic
hybridization between layers was 
not included in the model. The simplest extension that takes interlayer
hybridization into account includes a tight binding hopping element between
$\pi$ orbitals at Carbon atoms which are nearest neighbors in adjacent
layers. The introduction of this term changes substantially the electronic
structure of bulk 
graphite\cite{GNP06}, and also of systems containing few graphene
layers\cite{MF06,GNP06}.

The charge distribution in systems under an applied field must be calculated
self consistently. Such a calculation, using the interlayer hopping model
described above,  has been carried out for a graphene
bilayer\cite{MC06}. In the following, we calculate the charge
distribution in graphene stacks of arbitrary width, and in semi infinite
graphite. We calculate the electrostatic potential self consistently, and
assume that the induced charge can be obtained using linear response
theory. These assumptions amount to the Random Phase Approximation, applied
to the model mentioned earlier. A similar calculation for a bilayer
can be found in\cite{MC06}, and it is in reasonable agreement with more
involved numerical calculations.

We use, as a starting point, the calculations of the unpperturbed electronic
structure of finite graphene stacks, and semiinfinite graphite, reported
in\cite{GNP06}. We do not study the effects of other interlayer hoppings,
disorder, or other 
effects related to interactions. We also do not consider deviations from  
Bernal stacking, $\{1212 \cdots\}$, which can alter the electronic structure
at the Fermi level\cite{GNP06}.

The following section presents the model to be studied. We discuss then screening
in semiinfinite graphite, and we
analyze a finite graphene
stack next. Section \ref{conclusions} presents the main conclusions of our
work. It also contains a discussion of the limits of validity
of the results presented here, and their relation to previous work.
\section{The model.}
\subsection{Electrostatic effects.}
\begin{figure}
\begin{center}
\includegraphics*[width=6cm]{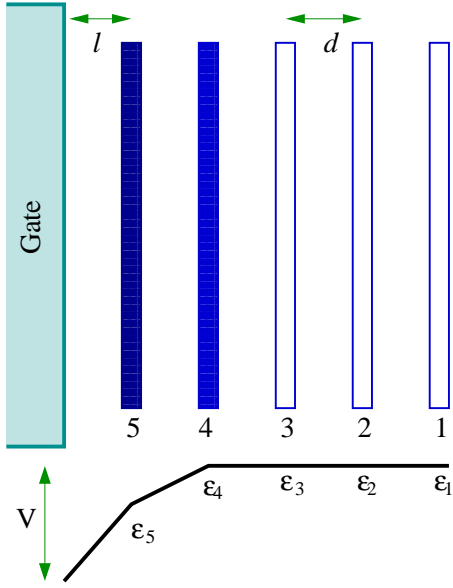}
\caption{(Color online). Sketch of the system studied in the text. 
The two layers closest to
  the gate are charged, and lead to the electrostatic potential shown in the
  lower part.}
\label{sketch}
\end{center}
\end{figure}

We analyze the charge distribution at the surface layers of a system with
many graphene layers coupled electrostatically to an external gate. The system
is schematically shown in Fig.[\ref{sketch}]. 
A potential $V$ is applied between the gate and the graphene stack. An
electric field, ${\cal E} = V / l$, where $l$ is the distance between the
gate and the stack. The electric potential beyond the gate is assumed to be
zero, so that the voltage at the gate is $-V$.
Using Gauss law we can write the total charge density per unit area in the
stack, $n$ as $e n = 4 \pi {\cal E}$.  This charge is distributed
among the layers, $n_1 , n_2 , n_3 , \cdots$ where label 1 stands for the
outermost layer.
The electric field in the region between layers 1 and
2 is ${\cal E}_{1-2} = 4 \pi e n_1 / d$, where $d$ is the interlayer
distance. This field detrmines the potential in layer 2. Extrapolating this
procedure to all layers, we find that the electrostatic potential in layer
$i$ satisfies:
\begin{equation}
\epsilon_i = \epsilon_{i-1} + 4\pi e^2 d \sum_{k=1}^{i-1} n_k 
\end{equation}
or, alternatively:
\begin{equation}
\epsilon_{i-1} - 2 \epsilon_i + \epsilon_{i+1} 
= 4 \pi e^2 d n_{i}
\label{potential}
\end{equation}
We include the possible effects of
a finite dielectric constant, $\epsilon_0$, into the definition of the
electrostatic charge, $e^2$.

\subsection{Linear response approximation.}
We analyze the system within the Hartree approximation, and we assume that the
induced charge can be well approximated using by linear response theory 
(Random Phase Approximation). 

\begin{figure}
\begin{center}
\includegraphics*[width=6cm]{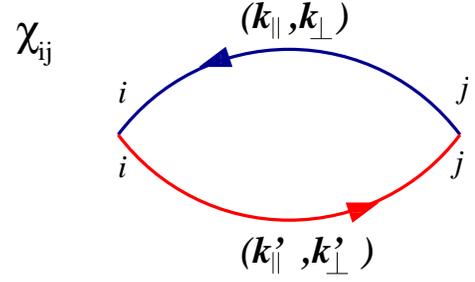}
\caption{(Color online. Charge susceptibility required in order to calculate the induced
  charge. }
\label{diagram}
\end{center}
\end{figure}

The induced electron density can be written, using linear response theory,
as:
\begin{equation}
n_i = \sum_j \chi_{i,j} \epsilon_j
\label{charge}
\end{equation}
where $\chi_{ij}$ is a static susceptibility which describes the charge
density induced at layer $i$ when the potential at layer $j$ is
$\epsilon_j$. The calculation of these susceptibilities is given by diagrams
like the one shown in Fig.[\ref{diagram}]. Their value is:

\begin{equation}
\chi_{ij} = \sum_{{\vk},{\vk}',\kp,\kpp} \frac{{\alphai}^* {\alphaip}^*
  \alphaj \alphajp}{\epsilon_{{\vk} , \kp} - \epsilon_{{\vk}',\kpp}}
\label{susc}
\end{equation}

where the intermediate empty and occupied states are labelled by their
moments, ${\vk}, \kp , {\vk}', \kpp$, and /i) and $j$ are layer indices.
The quantities $\alphai , \alphaj , \alphaip$ and $\alphajp$ in
eq.(\ref{susc}) are the amplitudes of the wavefunctions on layers $i$ and $j$
respectively. 
 We will only consider charge distributions 
which are homogeneous in the directions parallel to the layers, so that the
parallel momentum transfer in the diagram in Fig.[\ref{diagram}] vanishes, ${\bf
  \vec{q}} = {\vk} - {\vk}' = 0$.

\subsection{Electronic structure.}
We describe the electronic levels of the graphene stack by means of a tight
binding model using the $\pi$ orbitals at each C atom. We consider a hopping
between orbitals in atoms which are nearest neighbors within a given layer,
$t = 2.7$eV, and a hopping between atoms in adjacent layers which are also
nearest neighbors, $\tp = 0.3$eV. We study mostly the Bernal stacking, so
that the hopping term connects half of the orbitals in a given layer with half
of the orbitals in the two nearest layers. At long wavelengths and near half
filling the in plane dispersion is well approximated by the continuum Dirac
equation, described in terms of the Fermi velocity, $\vf = 3 t a / 2$, where
$a = 1.4$\AA is the distance between Carbon atoms. Details of the model, and of the
band structure for stacks with different number of layers and stacking
order are given in\cite{GNP06}.

 The system has 
translational invariance in the direction parallel to the layers, so that the
parallel momentum, ${\vk} \equiv ( k_x , k_y )$ is conserved. If the
stack is infinite, the momentum normal to the layers, $k_\perp$, is also
conserved. 

In an infinite stack with only nearest neighbor coupling between layers, all
layers are equivalent (in general, a description of the Bernal stacking 
requires two inequivalent layers). Then, 
hamiltonian for each momentum decouples in a set of $2 \times 2$ matrices, whose entries
correspond to Bloch states defined in the two inequivalent sublattices of
each layer. For a given corner of the Brillouin zone, the hamiltoonian can be
written as:
\begin{equation}
{\cal H}_{\vk , \kp} \equiv \left( \begin{array}{cc} 2 \tp \cos ( \kp d )
    &\vf ( k_x + i k_y ) \\ \vf ( k_x - i k_y ) &0 \end{array} \right) 
\label{hamil}
\end{equation}
The diagonal terms in eq.(\ref{hamil}) are determined by the interlayer
hopping, so that ${\cal H}_{22} = 0$, as one of the sublattices is decoupled
from the neighboring layers. The hamiltonian in eq.(\ref{hamil}) reduces to
the Dirac equation for $\tp = 0$. 

The low energy eigenenergies can be approximated as:
\begin{equation}
\epsilon_{{\vk} \kp} \approx \pm \frac{\vf^2 \mk^2}{2 \tp \cos ( \kp d )}
\label{ener_quad}
\end{equation} 
This approximation fails for $\kp d \sim \pi / 2$, where tha bands show a
linear dependence on $\vk$.

The allowed momenta in a finite stack with $N$ layers are quantized, $k_\perp^n
 = ( n \pi ) / [d (
N+1 )]$\cite{GNP06}. In addition,
the amplitude of the wavefunctions go to zero as $\sin ( i \kp c )$ at 
layer $i$ from the surface. The electronic
wavefunctions in a finite stack are described
 in Appendix \ref{wavefunction_app}.

\subsection{Calculation of the charge susceptibility.}
The response of the electrons at both inequivalent corners of the Brillouin
Zone is the same, so that we need to calculate the polarizability at
one $K$ point. 

\begin{figure}
\begin{center}
\includegraphics*[width=7cm]{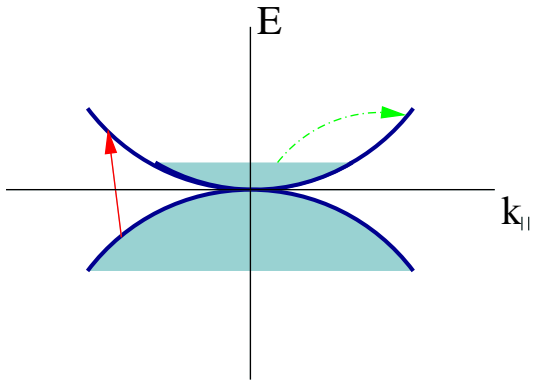}
\caption{(Color online). Sketch of the inter- and intraband transitions whose separate
  contributions to the charge susceptibility is analyzed in the text.}
\label{transitions_screening_fig}
\end{center}
\end{figure}

The charge susceptibility includes contributions from transitions between the
 valence and conduction bands, interband transitions, and trensitions within
 the conduction band, intraband transitions, as schematically depicted in
 Fig.[\ref{transitions_screening_fig}]. 

We consider first the interband transitions, involving
 an occcupied state in the valence band and an empty state in the conduction
 band. We neglect the contribution from the states with
$\kp d \sim \pi / 2$, and we use the approximate dispersion relation given in
 eq.(\ref{ener_quad}). Then, the susceptibilities in eq.(\ref{susc}) can be
written an integral over ${\vk}$, and $\kp$ and $\kpp$. The electronic states
 in a semiinfinite stack can also be characterized by a perpendicular
 momentum, $\kp$, although the corresponding wavefunctions are no longer
 running waves. Near the surface of
 a semiinfinite stack, the amplitudes in eq.(\ref{susc}) are:
\begin{equation}
\alphai = C \sin ( i \kp d )
\label{amplitude}
\end{equation}
where $C$ is a normalization constant. Note that, although one needs in
principle to define the amplitude as a two component spinor in each layer,
the low energy states considered here, eq.(\ref{ener_quad}), have vanishing
amplitude on the sublattice connected by the interlayer hopping $\tp$\cite{GNP06}.

\begin{figure}
\begin{center}
\includegraphics*[width=7cm,angle=-90]{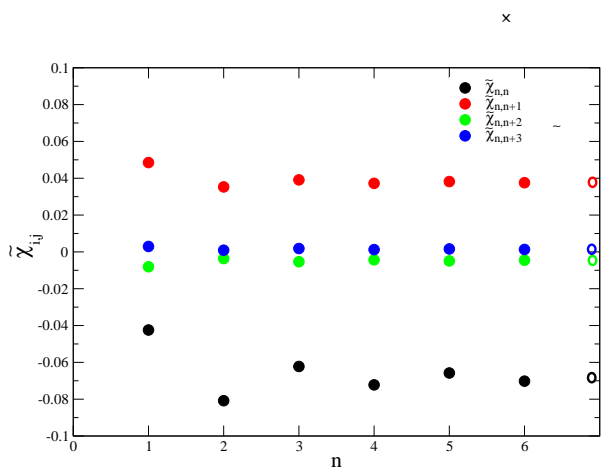}
\caption{(Color online). Value of the prefactor in the susceptibilities 
near the boundary of
  a stack of graphene planes, see eq.(\protect{\ref{susc2}}). The
  corresponding bulk values are plotted as open circles on the right.}
\label{susc_fig}
\end{center}
\end{figure}

Using eq.(\ref{amplitude}), we finally obtain:
\begin{widetext}
\begin{eqnarray}
\chiinter_{ij} &\approx &\frac{8}{\pi^3} \frac{\tp}{\vf^2}
\int_{ - \pi / 2}^{\pi / 2}
d \phi \int_{\pi / 2}^{( 3 \pi ) / 2} 
d \phi'
\frac{\sin ( i \phi ) \sin ( j \phi
  ) \sin ( i \phi' ) \sin ( j \phi') \cos ( \phi ) \cos ( \phi' )}{ [
\cos ( \phi ) + \cos ( \phi' ) ]} \int_0^{k_c} 
\frac{d k}{k} \nonumber \\ &= &\frac{4}{\pi} \frac{\tp}{\vf^2}
\tilde{\chi}_{ij} \log \left( \frac{\tp}{\eo} \right)
\label{susc2}
\end{eqnarray}
where $\eo$ is a low energy cutoff to be specified later, and we write $\phi
= \kp d$ and $\phi' = \kpp d$. In a finite
stack, these integrals over $\kp$ and $\kpp$  must be replaced by sums over the
quantized momenta, see section[\ref{finite_stack_section}].
The prefactor $\chitinter_{ij}$ in eq.(\ref{susc2}) is defined as:
\begin{equation}
\chitinter_{ij} = \frac{1}{\pi^2} \int_{-\pi / 2}^{\pi / 2} d \phi
\int_{\pi/2}^{3 \pi / 2} d \phi' \times
 \frac{\sin (i \phi) \sin(j \phi)
\sin(i \phi') \sin(j \phi') \cos(\phi) \cos(\phi')}{\cos(\phi) -\cos(\phi')} 
\label{susc3}
\end{equation} 
The values of $\chiinter_{ij}$ near the boundary of the stack are plotted
in Fig.[\ref{susc_fig}] The limiting bulk values:
\begin{equation}
\chitinter_{m+n,m} = \chitinter_{n} =
\frac{1}{4 \pi^2} \int_{-\pi / 2}^{\pi / 2} d \phi
\int_{\pi/2}^{3 \pi / 2} d \phi' \frac{\cos  [ n ( \phi - \phi' )] 
\cos(\phi) \cos(\phi')}{\cos(\phi) -\cos(\phi')}
\label{bulk_n}
\end{equation}
are also shown.
\section{Results.}
\subsection{Undoped semiinfinite stack.}
In a semiinfinite stack, sufficiently far from the surface, so that the
susceptibilities $\chi_{ij}$ have converged towards their bulk values, 
eqs.(\ref{potential}) 
and (\ref{charge}) admit the solution:
\begin{eqnarray}
\epsilon_i & = &\epsilon_0 e^{- \kappa} \nonumber \\
n_i &= &n_0  e^{- \kappa}
\label{solution}
\end{eqnarray}
with:
\begin{equation}
e^{\kappa} + e^{-\kappa} - 2 = 4 \pi e^2 d \sum_{n = - \infty}^{\infty} e^{n
  \kappa} \chiinter_n = 
\frac{8 e^2 d \tp}{\pi \vf^2} \log \left( \frac{\tp}{\epsilon_0}
  \right) \left\{ \frac{1 + \cosh ( \kappa )}{2 \sinh ( \kappa / 2 )} \arctan
  \left[ \sinh \left( \frac{\kappa}{2} \right) \right] - 1 \right\}
\label{bulk}
\end{equation}
\end{widetext}
Details of the steps involved in the derivation of eq.(\ref{bulk}) are given
in  Appendix \ref{susc_app}.

In metallic systems and for $\kappa \rightarrow 1$, 
the right hand side in eq.(\ref{bulk}) is equal to
$\lim_{\kp \rightarrow 0} \chi ( \kp , {\bf \vec{k}}_\parallel = 0 ) = D (
\ef )$, where $D ( \ef )$ is the density of states at the Fermi level in a
given layer. Using this result, we obtain the Thomas-Fermi approximation, 
$\kappa^2 =  4 \pi e^2 d D ( \ef )$ in the limit $\kappa \rightarrow
0$. 

For the model for undoped graphite studied here, the r.h.s. of
eq.(\ref{bulk}) vanishes as $\kappa \rightarrow 0$, and ordinary screening
does not take place. 

\begin{figure}
\begin{center}
\includegraphics*[width=7cm,angle=-90]{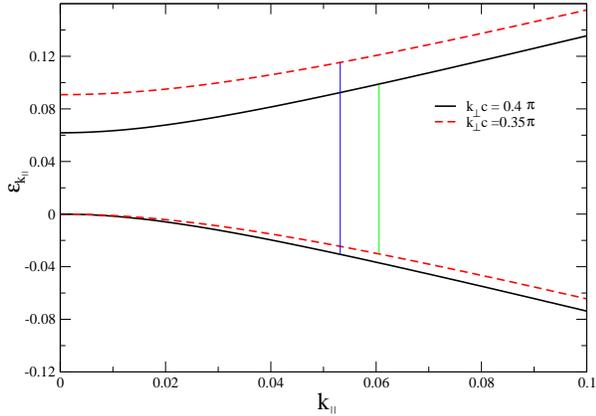}
\caption{(Color online). Transitions between the valence and conduction bands
states for $( k_\perp -
  k_\perp' ) c = 0.05 \pi$ and $k_\perp c = 0.4 \pi$. The apameters used are
  $\vf = 1$ and $\tp = 0.1$.}
\label{transition_fig}
\end{center}
\end{figure}

The contribution of the interband transitions to the
charge susceptibility is not enough to give a finite charge compressibility
at long wavelengths, leading to a behavior reminiscent of that of 
an insulator.  
 Note that, at zero doping, 
the transitions between occupied and empty states
in the limit $\phi - \phi'\rightarrow 0$ require a finite energy of order
$\tp \cos ( \phi )$, except for
$\phi \approx \phi' \approx \pi / 2$. These transitions have vanishing weight
in the integral which gives the charge compressibility (see
Fig.[\ref{transition_fig}]). Hence, the charge susceptibility tends to zero at
zero energy and $k_\perp d \rightarrow 0$. 

Note that, on the other hand, the staggered
spin and charge susceptibilities, $k_\perp d \ne 0$, diverge
logarithmically at $\omega \rightarrow 0$\cite{NCPG06}. This divergence is
maximal for $\kp d = \pi$.

Eq.(\ref{bulk}) admits only solutions with $\kappa > 0$ if:
\begin{equation}
\frac{4 e^2 d \tp}{\pi \vf^2} \log \left( \frac{\tp}{\epsilon_0} \right)
\ge 6
\label{critical}
\end{equation}
The scale $\epsilon_0$ in eq.(\ref{susc2}) 
determines the region where the model gives a valid
description of a graphene stack. At zero temperature, it will be determined by
interlayer hoppings not considered here, disorder, and lifetime broadening
due to interaction effects. At finite temperatures, the logarithmic
divergence in eq.(\ref{susc2}) is cutoff by thermal interband excitations, st
that $\epsilon_0 \approx T$. Hence, the screening properties of the model
depend on temperature.

At the surface, the screening by interband transitions is modified, as the
relative strength of the different subbands, as function of $\kp$, is
modified. The contribution to the real space matrix elements of the
susceptibility are given by:
\begin{equation}
  \chiintra_{ij} = \frac{8 \tp}{\pi^2 \vf^2}
\int_{- \pi / 2}^{\pi / 2} d
  \phi \sin ( \phi i )^2 \sin ( \phi  j )^2 \cos ( \phi )
\label{chi_intra}
\end{equation}

At finite dopings the density of states of each graphene plane in the model
studied here is:
\begin{equation}
D ( \ef ) = \frac{4 \tp}{\pi^2 \vf^2}
\label{dos}
\end{equation}
independent of the carrier concentration. Intraband transitions modify the
charge susceptibility as$\kp \rightarrow 0$. The charge compressibility
becomes finite, leading to a  bulk 
screening length : 
\begin{equation}
\kappa^2 = \frac{16 e^2 d \tp}{\pi \vf^2}
\label{kappa}
\end{equation}

In the following, we use the parameters:
\begin{eqnarray}
\frac{e^2}{\vf} &\approx &1 \nonumber \\
\frac{\tp d}{\vf} &\approx &\frac{1}{5}
\label{parameters}
\end{eqnarray}
Using these values, eq.(\ref{kappa}) gives a screening lengt due to bulk
intraband transitions of  about $N \sim 2$ layers. 
The screening at the surface, however, is modified and reduced, as shown in
eq.(\ref{chi_intra}). 

\begin{figure}
\begin{center}
\includegraphics*[width=7cm,angle=-90]{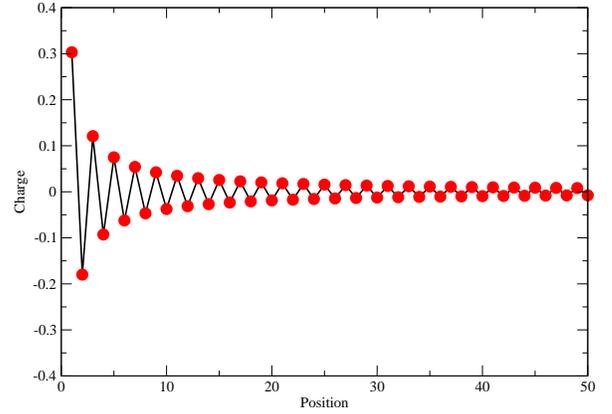}
\caption{(Color online). Charge at the surface of a semiinfinite stack, normalized to
  one. Only interband transitions are included in the calculation.}
\label{charge_surface_fig}
\end{center}
\end{figure}

 Numerical results, including the full dependence on position of
 $\tilde{\chi}_{m,m+n}$  are shown in
Fig.[\ref{charge_surface_fig}]. We have chosen the parameter $\epsilon_0$
 such that $\log ( \tp / \epsilon_0 ) = 6$. 
The charge oscillates with every second
layer. This result is consistent with the logarithmic divergence of the
charge and spin susceptibility for $k_\perp d = \pi$. The induced charge is
 reduced to less than $10 \%$ of the charge in the surface layer in about $N
 \sim 3-5$ layers.
 
\subsection{ Finite stacks.}
\label{finite_stack_section}
The eqs.(\ref{potential}) and (\ref{charge}) can be solved in a
straightforward way for a system with a few layers. The discrete equivalent
to the reduced susceptibilities in eq.(\ref{susc3}) can be defined as:
\begin{eqnarray}
\chitinter_{ij} &= &\sum_{n=1}^N \sum_{n'=1}^N \sin( k_\perp^n d i )
  \sin ( k_\perp^n d j ) \sin ( k_\perp^{n'} d i ) \sin ( k_\perp^{n'} d j ) 
\times \nonumber \\ &\times &\frac{
  \cos ( k_\perp^n d ) \cos ( k_\perp^{n'} d )}{C_{k_\perp^n} C_{k_\perp^{n'}}
  \left[ \cos ( k_\perp^n d ) - \cos (
  k_\perp^{n'} d ) \right]}a
\label{susc_finite}
\end{eqnarray}

\begin{figure}
\begin{center}
\includegraphics*[width=7cm,angle=-90]{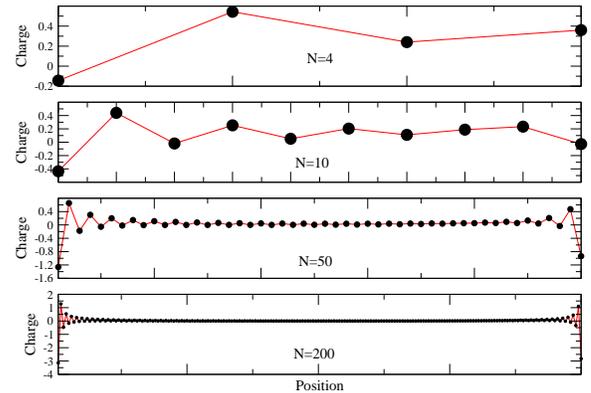}
\caption{(Color online). Charge distribution in systems with 4, 10, 50 and 200 graphene
  layers, normalized to one.}
\label{charge_fig}
\end{center}
\end{figure}

where $C_{k_\perp^n}$ is a normalization constant:
\begin{equation}
C_{k_\perp^n} = \sum_{l=1}^N \sin^2 ( k_\perp^n c l )
\end{equation}
A finite bilayer can be charged, so that the Fermi energy does not need to
lie exactly
between the valence and conduction bands. Then, we have to add to the
susceptibility a contribution from intraband transitions at ${\bf
  \vec{k}}_\parallel \rightarrow 0$. The contribution of each subband is
determined by its contribution to the total density of states, and it is
independent of the position of the Fermi level. 
Hence, intraband transitions lead to:
\begin{equation}
\chiintra_{ij} = \frac{2 \tp}{\pi^2 \vf^2} \sum_n \frac{\sin^2 (
    k_\perp^n d i ) \sin^2 ( k_\perp^n d j ) \left| \cos \left(
    k_\perp^n d \right) \right|}{C_{k_\perp^n}^2}
\end{equation}
The charge distribution is given by the equations:
\begin{eqnarray}
\epsilon_i &= &\epsilon_{i-1} + 4 \pi e^2 d \sum_{k=1}^{i-1} n_k \nonumber \\
n_i &= &n_0 + \sum_{j=1}^N \left( \chi_{ij}^{\rm inter} + \chi_{ij}^{\rm intra}
\right) \epsilon_j
\end{eqnarray}
where $n_0$ is a constant which fixes the total density, in turn detrmined by
the applied field between the gate and the stack. 

The diagonal intraband susceptibilities near the surface of a
semiinfinite stack satisfy:
\begin{eqnarray}
\chitintra_{nn} &\propto &\int_{- \pi / 2}^{\pi / 2} d \phi 
\sin^4 ( n \phi ) \cos (
\phi ) = \nonumber \\ 
&= &- \frac{1}{4 ( 16 n^2 - 1 )} + \frac{(-1)^n}{2 n^2 - 1} +
\frac{3}{4}
\end{eqnarray}
showing even-odd oscillations as function of the distance to the surface,
$n$, as well as a slow convergence to the bulk limit, $n \rightarrow \infty$.

If we neglect the intraband susceptibility, we find for a bilayer:
\begin{equation}
\epsilon_1 - \epsilon_2 = \frac{2 \pi e^2 d n_0}{1 - 2 \pi e^2 d \left(
    \chi_{11} - \chi_{12} \right) }
\label{bilayer}
\end{equation}
where $n$ is the total carrier density, and $\chi_{11}$ and $\chi_{12}$ are
the bilayer interband susceptibilities, defined using
eqs.(\ref{susc_finite}) and (\ref{susc2}). 
By choosing the low energy cutoff, $\eo$  such that
$\eo = \ef = ( \pi \vf^2 n_0 ) / \tp$, 
we recover the results in\cite{MC06} to lowest
order in $n_0$.

Examples of the charge distribution in doped systems with different number of
layers are shown in Fig.[\ref{charge_fig}]. We find that: 

i) The charge
distribution is rather homogeneous in narrow stacks, and it becomes
concentrated at the surfaces for stacks wider than the screening
length. 

ii) There are 
oscillations as function of the distance to the
surface in wide stacks, as in the case of a semiinfinite stack. These
oscillations can lead to charge with different sign in neighboring layers. 

\section{Conclusions.}
\label{conclusions}
We have calculated the charge distribution in stacks of graphene layers in an
applied field. We describe the electronic structure by a tight binding model
for the $\pi$ orbitals, which includes hopping between sites which are
nearest neighbors in adjacent layers. The self consistent electrostatic
potential is obtained assuming linear response theory, so that our
approximations amount to the Random Phase Approximation.

The electronic structure of these systems shows a valence and conduction
bands with parabolic dispersion as function of the parallel momentum, for all
values of the perpendicular momentum. These bands touch at zero energy, which
defines the chemical potential in the undoped case. Finite and infinite
stacks are gapless.

The charge susceptibility can be written as a sum of intra- and
interband contributions. In a clean, neutral  system only interband
transitions contribute. In this case, an infinite stack of layers shows
non metallic screening, as the long wavelength charge susceptibility
vanishes. 

The charge susceptibility for finite momenta
perpendicular to the layers shows a logarithmic divergence at low
energies. This divergence is maximal for a wavevector inversely proportional
to the distance between the layers, leading to charge oscillations. In
addition, the logarithmic divergence at finite wavevectors implies that the
screening properties of undoped systems can show a depedence on the frequency
at which they are probed, or on temperature.

The charge distribution induced at a surface can be
extended over many layers, and it shows a decaying modulation with a period
equal to the distance between the layers.

In doped samples, the long wavelength charge polarizability is finite, and it
is dominated by
intraband transitions. We find that an external field is screened within 3-5
layers  from the surface. The screening length is independent of carrier
concentration. The interband transitions lead to oscillations in the induced
charge, as in the undoped case. As these fluctuations depend logarithmically
on a low energy cutoff, they can also show a dependence on temperature or
frequency. 

The screening length in doped stacks obtained here depends on the values of
the parameters given in (\ref{parameters}). Their bulk values are not known
with precision\cite{BCP88}, and it is possible that some of them,
like the effective electric charge or the interlayer hopping, change
near an interface. Hence, the value of 3-4 layers over which the charge is
delocalized in doped samples is only approximate.

Our calculation does not take into account effects such as next nearest
neighbor hoppings, disorder, or deviations from linear reponse. The existence
of other hoppings, or disorder, will define a low energy scale below which
the results will no longer be valid. 

Deviations from linear response theory
depend on the strength of the induced electrostatic potential with respect to
the parameters which define the band structure, the smaller of which is the
interlayer hopping, $\tp$, which we have assumed 
to be $\tp  \sim 0.1 - 0.3$eV. 
Typical differences in electrostatic potentials
between adjacent layers are given by $\epsilon_i - \epsilon_{i-1} \approx e^2
d n$, where $n$ is the induced charge per unit area. For $n \sim 10^{11} - 
10^{12}$ cm$^{-2}$, we obtain $\epsilon_i - \epsilon_{i-1} 
\sim 10^{-3} - 10^{-2}$eV, so that the assumption of
linear response is probably valid. Finally, we have only studied the Bernal
stacking, $1212 \cdots$. Regions of rhombohedral stacking reduce the density
of states near the Fermi level\cite{GNP06}, and they 
will decrease the screening in the system. 

It is interesting to consider the similarities and differences of this work 
with the related calculation in\cite{VF71}. 
In this reference, it was assumed that a
low density two dimensional electron gas existed in each layer, and that the
electrons could not move between different layers. This approximation is
equivalent to consider only the diagonal susceptibility, $\chi_{nn}$, in
eq.(\ref{susc2}). For the 2DEG, only intraband
transitions exist, leading to $\chi_{nn} = D ( \ef ) = n_{\rm v} \times 
(m^* / (\pi \hbar^2 )$, where $n_{\rm v}$ is the number of valleys. 
This assumption leads to metallic screening, with a
decay of the electrostatic potential into the bulk, eq.(\ref{bulk}), given by
$\kappa^2 \approx ( e^2 d n_{\rm v} m^* ) /  (\pi \hbar^2 )$. In our case,
the existence of a finite density of states near the Fermi level is a
consequence of the finite interlayer hopping, $\tp$, 
as the electronic structure
reduces to a stack of decoupled Dirac equations in the absence of
hopping. The charge oscillations near a surface are directly related to the
quantum coherence between stacks induced by the interlayer hopping. Hence,
they cannot be obtained in the model used in\cite{VF71}.

\section{Acknowledgements.}
I am thankful to A. Castro Neto for many helpful discussions. Financial
support from the Ministerio de Educaci\'on y Ciencia (Spain) through grant
no. FIS2005-05478-C02-01, the European Union, through contract 12881, and and
the program CITECNOMIK 
(C. A. Madrid),  ref. CM2006-S-0505-ESP-0337 is gratefully acknowledged.
\appendix
\section{Electronic wavefunctions in finite stacks.}
\label{wavefunction_app}
We define the amplitude of the wavefunction with parallel momentum $\vk$,
perpendicular momentum $\kp^n$ at layer $M$ as a two component spinor:
\begin{equation}
\Psi_{ \vk , \kp^n , M } \equiv 
\left( \begin{array}{c} \aAM \\ \aBM \end{array}
\right) 
\end{equation}
where $\aAM$ and $\aBM$ refer to the amplitudes on the $A$ sublattice, whose
atoms in different layers are nearest neighbors, and the $B$ sublattice,
whose atoms in different layers are not
connected. In order to satisfy the open boundary conditions at the surfaces
of the stack, these amplitudes must be of the form:
\begin{equation}
\left( \begin{array}{c} \aAM \\ \aBM \end{array}
\right) \equiv \left( \begin{array}{c} \aA \\ \aB \end{array}
\right) \sin \left( \kp^n c M \right)
\end{equation}
and:
\begin{widetext}
\begin{equation}
\left( \begin{array}{cc} 2 \tp \cos ( \kp^n c ) &\vf ( k_x + i k_y ) \\ \vf (
    k_x - k_y ) & 0 \end{array} \right) \left( \begin{array}{c} \aA \\ \aB 
\end{array} \right) = \epsilon_{{\vk} , \kp} 
\left( \begin{array}{c} \aA \\ \aB 
\end{array} \right)
\label{hamil_app}
\end{equation}
The low energy eigenvalues, $| \epsilon_{{\vk},\kp^n} | \ll \tp$ are given by:
\begin{equation}
\epsilon_{{\vk},\kp^n} = \pm \tp 
\cos ( \kp^n c ) \mp \sqrt{\tp^2 \cos^2 ( \kp c
  ) + \vf^2 \mk^2} \approx \mp \frac{\vf^2 \mk^2}{2 \tp \cos ( \kp^n c )}
\end{equation}
\end{widetext}
within this low energy approximation, eq.(\ref{hamil_app}) implies that:
\begin{equation}
\left( \begin{array}{c} \aA \\ \aB 
\end{array} \right) \approx \left( \begin{array}{c} C^{1/2}_{\kp^n}
 \\ 0 \end{array} \right) 
\end{equation}
where $C^{1/2}_{\kp^n}$ is a normalization constant.
\section{Calculation of the bulk susceptibility.}
\label{susc_app}
We derive the right hand side of eq.(\ref{bulk}) using the expressions in
eq.(\ref{bulk_n}). The dependence on the layer index $n$ in eq.(\ref{bulk_n})
is through the factor $\cos [ n ( \phi - \phi' ) ]$. In order to regularize
the summations over $n$, we adda small decaying factor, $\epsilon$:
\begin{equation}
\cos [ n ( \phi - \phi' ) ] \rightarrow e^{- \epsilon | n |} \cos [ n ( \phi -
\phi' ) ] 
\end{equation}
For $\kappa = 0$ in eq.(\ref{bulk}) the $n$ dependent part of the sum gives:
\begin{widetext}
\begin{equation}
\sum_{n=-\infty}^{\infty}  e^{- \epsilon | n |} e^{i n ( \phi_0 +
\phi - \phi' ) } = \frac{\sinh ( \epsilon)}
{\cosh ( \epsilon ) - \cos ( \phi_0 + \phi - \phi' )} 
\xrightarrow{\epsilon \rightarrow 0}
2 \pi \delta ( \phi_0 + \phi - \phi' )
\label{shift}
\end{equation}
where, for convenience, we have also incuded a shift, $\phi_0$.
Using this result, we also obtain:
\begin{eqnarray}
{\cal F} ( \phi_0 ) &= &
\int_{-\pi/2}^{\pi/2} d \phi \int_{\pi/2}^{3 \pi / 2} d \phi'
\sum_{n=-\infty}^{\infty} \cos [ n ( \phi_0 + \phi - \phi' ) ] \frac{\cos (
  \phi ) \cos ( \phi' )}{\cos ( \phi ) - \cos ( \phi' )} 
\int_{-\pi/2}^{\pi/2} d \phi \frac{\cos (
  \phi ) \cos ( \phi_0 + \phi)}{\cos ( \phi_0 + \phi ) - \cos ( \phi )} = 
\nonumber \\ &= &\frac{1 + \cos ( \phi_0 )}{4 \sin ( \phi_0 / 2 )} \log \left[
  \frac{1 + \sin ( \phi_0 / 2 )}{1 - \sin ( \phi_0 / 2 )} \right] - 1 
\end{eqnarray}
\end{widetext}
The summation for $\kappa \ne 0$ in eq.(\ref{bulk}) is formally equivalent to
the replacement $\phi_0 \rightarrow i \kappa$. Making this substitution, we
find: 
\begin{equation}
{\cal F} ( i \kappa ) = \frac{1 + \cosh ( \kappa )}{2 \sinh ( \kappa / 2 )}
\arctan \left[ \sinh \left( \frac{\kappa}{2} \right) \right] - 1
\end{equation}
For $\kappa \rightarrow 0$ we find:
\begin{equation}
{\cal F} ( i \kappa ) \approx \frac{\kappa^2}{6} + \frac{\kappa^4}{180} +
\cdots 
\end{equation}
\bibliography{screening_surface}
\end{document}